\documentclass[aps,prc,preprint,showpacs,superscriptaddress,floatfix]{revtex4-1}
\usepackage[dvips]{color} 
\usepackage{graphicx} 
\usepackage{bm}

\begin{document}

\title{Anti-cluster Decay and Anti-alpha Decay of Antimatter nuclei}

\author{D. N. Poenaru}
\email[]{poenaru@fias.uni-frankfurt.de}
\affiliation{
Horia Hulubei National Institute of Physics and Nuclear
Engineering (IFIN-HH), \\P.O. Box MG-6, RO-077125 Bucharest-Magurele, Romania}
\affiliation{Frankfurt Institute for Advanced Studies (FIAS),
Ruth-Moufang-Str. 1, 60438 Frankfurt am Main, Germany}
\author{R. A. Gherghescu}
\affiliation{
Horia Hulubei National Institute of Physics and Nuclear
Engineering (IFIN-HH), \\P.O. Box MG-6, RO-077125 Bucharest-Magurele, Romania}
\author{W. Greiner}
\affiliation{Frankfurt Institute for Advanced Studies (FIAS),
Ruth-Moufang-Str. 1, 60438 Frankfurt am Main, Germany}

\date{ }

\begin{abstract}

A broad extension of periodic system into the sector of antimatter could be
possible sometimes in a remote future.  We expect that anti-alpha
spontaneous emission from an antimatter nucleus will have the same Q-value
and half-life as alpha emission from the corresponding mirror nucleus.  This
is the consequence of the invariance of binding energy as well as of the
surface and Coulomb energy when passing from matter to antimatter nuclei
with the the same mass number and the same atomic number.  The Q-values and
half-lives of all measured up to now 27 cluster radioactivities are given
together with Q-values and half-lives of the most important competitor ---
$\alpha$~decay.  The lightest anti-alpha emitter, $^8\bar{Be}$, will have a
very short half-life of about $81.9\cdot 10^{-18}$~s.

\end{abstract}


\maketitle

\section{Introduction}
\label{sec:1}


In 1928 Dirac 
predicted the existence of negative energy states of electrons
\cite{dir28prs} when he developed his famous relativistic wave equation for
massive fermions.  The antimatter character of these states became clear in
1933 after discovery of the positron (the antielectron) in cosmic radiation
by Anderson 
\cite{and33pr}.

Individual anti-particles are produced by particle accelerators and in some
types of radioactive decay.  Antiprotons ($\bar{p}$) \cite{cha55pr} were
observed in 1955 by Segr\`e and Chamberlain. 
The antineutron was discovered in proton-proton collisions at the Bevatron
(Lawrence Berkeley National Laboratory) by Cork et al.  in 1956
\cite{cor56pr}.  Antiprotons are produced at Fermilab for collider physics
operations in the Tevatron.  Other accelerators with complex projects for
antimatter physics are the Relativistic Heavy Ion Collider (RHIC) at
Brookhaven National Laboratory, LHC at CERN, and in the future Facility for
Antiproton and Ion Research (FAIR)'s high-energy storage ring in Darmstadt
\cite{and11plb}.

Until now it was established in all experiments that every antiparticle has
the same mass with its particle counterpart; they differ essentially by the
sign of electric charge.  Also every antinucleus has the same mass or
binding energy as its mirror nucleus \cite{clo09b}.

Anti-atoms are difficult to produce; the simplest one --- the antihydrogen
($\bar{H}$) was produced, cooled and confined \cite{qui14b} for about 1000~s
\cite{bau96pl,gab02prl,and10n,alp11np}.  At the beginning the Low Energy
Antiproton Ring (LEAR) at CERN was used.  This device decelerated the
antiprotons and stored them in a ring.  The antimatter helium-4 nucleus,
$^4\bar{He}$, or anti-$\alpha$, consists of two antiprotons and two
antineutrons (baryon number $B = -4$) \cite{sta11n}.  This is the heaviest
observed antinucleus to date.  It seems that the next one, antilithium, has
an extremely low production rate.  

It will be a long way to produce a rich diversity of more complex antinuclei
justifying a broad extension of periodic system into the sector of
antimatter and strangeness \cite{gre12jpcs}.  Nevertheless in this work we
try to understand whether their decay modes by anti-$\alpha$ and
anti-cluster spontaneous emission would differ from $\alpha$~decay and
cluster radioactivity \cite{enc95,p302bb10,p309prl11} of corresponding
mirror nuclei.


\section{Potential Barriers}

Let us assume that a binary decay mode (e.g. anti-alpha decay, anti-cluster
decay or spontaneous fission) of a parent anti-nucleus,
$^{A}\bar{Z}$, leads to an emitted anti-cluster, $^{A_e}\bar{Z_e}$, and a
daughter anti-nucleus, $^{A_d}\bar{Z_d}$:
\begin{equation}
^{A}\bar{Z} \ \rightarrow \ ^{A_d}\bar{Z_d} \ + \ ^{A_e}\bar{Z_e}
\end{equation}
with conservation of baryon numbers. Alternatively the subscript $d$ may be
denoted with $1$ and $e$ with $2$. By definition, the number of antiprotons
of an antinucleus is equal with the number of protons of the corresponding
nucleus. The same is true for the number of antineutrons and of neutrons.
Consequently, there is a good reason to assume that
every anti-cluster, $^{A_e}\bar{Z_e}$, will have the same binding energy as
the cluster $^{A_e}Z_e$, and similarly the binding energy of the daughter
anti-nucleus, $^{A_d}\bar{Z_d}$, will be identical with that of the
daughter, $^{A_d}Z_d$, and the binding energy of the parent anti-nucleus,
$^{A}\bar{Z}$, is identical with that of the parent $^{A}Z$. The released energy
\begin{equation}
Q = [M - (M_d + M_e)]c^2
\end{equation}
can be calculated using the last evaluation of experimental data for atomic
masses \cite{wan12cpc}. In this eq. $c$ is the light velocity, $M, M_d, M_e$
are the masses of parent, daughter and emitted nucleus. 

In general the ratio of $Z/A  \neq Z_d/A_d \neq Z_e/A_e$ meaning that the
three partners have different charge densities.  One can take into
consideration the difference in charge densities \cite{p80cpc80} by assuming
uniformity in each of the two fragments.  In this way the nuclear volume
$V=V_1 + V_2$ is divided in two parts, each of them being homogeneously
charged with a density
\begin{equation}
\rho _e ({\bf r}) = \left \{ \begin{array}{ll} \rho _{1e},& {\bf r}
\in V_1 \\ \nonumber
\rho _{2e},& {\bf r} \in V_2 \\    \end{array} \right .
\end{equation}

During the decay process from one parent to two fragments there is a
potential barrier which determines the metastability of any anti-nucleus. 
It is penetrated by quantum mechanical tunnelling as was shown by Gamow in
1928 for alpha~decay of nuclides \cite{gam28zp}.  

For cylindrical symmetry the simplest parametrization of the shape during
this process, with only one deformation parameter (the volume and the radius
of the emitted fragment are conserved), is that of two intersected spheres
assumed in the two-center shell model \cite{gre96bt}.  The radius of the
initial spherical anti-nucleus is $R_0=r_0A^{1/3}$ and the radii of the two
fragments are $R_e=r_0A_e^{1/3}$ and $R_d=r_0A_d^{1/3}$.  Within
Myers-Swiatecki's liquid drop model (LDM) \cite{mye66np} the radius constant
$r_0=1.2249$ fm and in the Yukawa-plus-Exponential model (Y+EM)
\cite{kra79pr} $r_0 = 1.16$ fm.  During the overlapping stage the separation
distance of the two fragments increases from an initial value $R_i=R_0-R_e$
to the touching point value $R_t=R_e + R_d$.  It is convenient to use the
deformation parameter $\xi=(R-R_i)/(R_t - R_i)$ equal to unity at the
touching point $R=R_t$.

We apply the macroscopic-microscopic method \cite{str67np} to calculate the
deformation energy, $E_{def}$, according to which a small shell and pairing
correction $\delta E$ is added to the macroscopic phenomenological model
deformation energy obtained by summing the surface and Coulomb energy due to
the strong and electrostatic forces:
\begin{equation}
E_{def}= (E_s - E_s^0) + (E_C - E_C^0)
\end{equation}
where $E_s^0=a_{20}A^{2/3} = a_s(1-\kappa_sI^2)A^{2/3}$ 
and $E_C^0=3e^2 Z^2 /(5r_0 A^{1/3}$ correspond to the spherical parent
with $a_s=17.9439$~MeV, $I=(N-Z)/A$ and $\kappa_s=1.7826$ within LDM.

The proton levels and neutron levels of a single particle shell model, e.g. 
two center shell model \cite{ghe03prc}, allowing to calculate \cite{str67np}
the shell and pairing correction, $\delta E$, are different because protons
are electrically charged.  In the same way for antinuclei the antiproton
levels should be different from antineutron levels but the antiproton levels
would be identical with proton levels and antineutron levels identical with
neutron levels.

\subsection{Strong interaction}
During the deformation from $R=R_i$ to $R=R_t$ the strong interaction is
responsible for the surface energy.  The strong force acts between
antinucleons in the same manner it acts between nucleons; the electric
charge doesn't play any role. For a number of
antinucleons equal to that of nucleons it will have the same effect. The
deformation dependent term is obtained by division with $E_s^0$:
\begin{equation} 
B_s=\frac{E_s}{E_s ^0} =
\frac{a_{21}}{a_{20}} B_{s1} + \frac{a_{22}}{a_{20}} B_{s2} 
\end{equation}
with $a_{21} \neq a_{22} \neq a_{20}$ taking into account the difference in
charge densities.  $B_{s1}$ and $B_{s2}$ are proportional with surface areas
of the fragments: 
\begin{equation} 
B_{s1}= \frac{d^2}{2} \int _{-1} ^{x_c}
\left [y^2+\frac{1}{4} \left ( \frac{dy^2}{dx}\right ) ^2 \right ]^{1/2} dx
\end{equation} 
\begin{equation} 
B_{s2}= \frac{d^2}{2} \int _{x_c} ^{1} \left
[y^2+\frac{1}{4} \left ( \frac{dy^2}{dx}\right ) ^2 \right ]^{1/2} dx
\end{equation} 
where $d=(z'' - z')/2R_0$ is the length of the deformed antinucleus divided
by the diameter of the spherical shape and $x_c$ is the position of
separation plane between fragments with -1, +1 intercepts on the symmetry
axis (surface equation $y = y(x)$ or $y_1 = y(x')$).

\subsection{Coulomb interaction}

We can see that not only the surface energy but also the Coulomb energy is
invariant when passing from matter to antimatter because in the following
general relationship \cite{dav75jcp} 
the charge density appears as a product of
$\rho_e({\bf r})\rho_e({\bf r_1})$:
\begin{equation}
E_c = \frac {1}{2} \int_{V_n} \int \frac {\rho _e ({\bf r})
\rho _e ({\bf r}_1) d^3 r d^3 r_1}{|{\bf r} - {\bf r}_1|}
\label{ecoul}
\end{equation}
See also the expression of $E^0_C$ above. 

For fragments with different charge densities by dividing with $E^0_C$ we
obtain
\begin{equation}
B_c =\frac{E_c}{E_c ^0} = \left ( \frac{\rho _{1e}}{\rho _{0e}}
\right ) ^2 B_{c1} + \frac{\rho _{1e} \rho _{2e}}{\rho _{0e} ^2}
B_{c12} + \left ( \frac{\rho _{2e}}{\rho _{0e}} \right ) ^2 B_{c2} 
\end{equation}
explicitly showing the electrostatic self-energies and the interaction of two
fragments.
For binary systems with different charge densities
and axially-symmetric shapes, we got 
\begin{equation}
B_{c1}= b_c \int _{-1} ^{x_c} dx \int _{-1} ^{x_c} dx' F(x, x') 
\end{equation}
\begin{equation}
B_{c12}= b_c \int _{-1} ^{x_c} dx \int _{x_c} ^{1} dx' F(x, x') 
\end{equation}
\begin{equation}
B_{c2}= b_c \int _{x_c} ^{1} dx \int _{x_c} ^{1} dx' F(x, x') 
\end{equation}
where $b_c = 5d^5 /8\pi$ and $d,x_c$ were defined in the previous
subsection.
The integrand is given by   
\begin{eqnarray} 
F(x,x')&=&\{ y y_1\frac{K-2D}{3}
\left [ 2(y^2+y_1^2)-(x-x')^2+\frac{3}{2}(x
-x')\left ( \frac{dy_1^2}{dx'}-\frac{dy^2}{dx} \right ) \right ]
\nonumber \\ 
 & & +K \left \{ \frac{y^2y_1^2}{3}+\left [y^2-
\frac{x-x'}{2}\frac{dy^2}{dx}
\right ] \left [y_1^2+\frac{x-x'}{2}\frac{dy_1^2}{dx'}\right ]
\right \} \} a_{\rho}^{-1} 
\end{eqnarray}  
where $D = (K - K')/k^2$; 
$K$ and $K'$ are the complete elliptic integrals of the first and second  
kind, respectively:
\begin{equation}   
K(k) = \int _0^{\pi /2}(1-k^2 {\sin}^2 t)^{-1/2} dt \; ; \;
K'(k) = \int _0^{\pi /2}(1-k^2 {\sin}^2 t)^{1/2} dt 
\end{equation}
and $a_{\rho} ^2 = (y+y_1)^2+(x-x')^2$, $k^2 = 4yy_1 /a_{\rho}^2$.  The
elliptic integrals are calculated by using the Chebyshev polynomial
approximation.  For $x = x'$ the function $F(x,x')$ is not determined.  In
this case, after removing the indetermination, we get $F(x,x')=4y^3/3$.

\subsection{Examples}

\begin{figure}[htb]
\includegraphics[width=14cm]{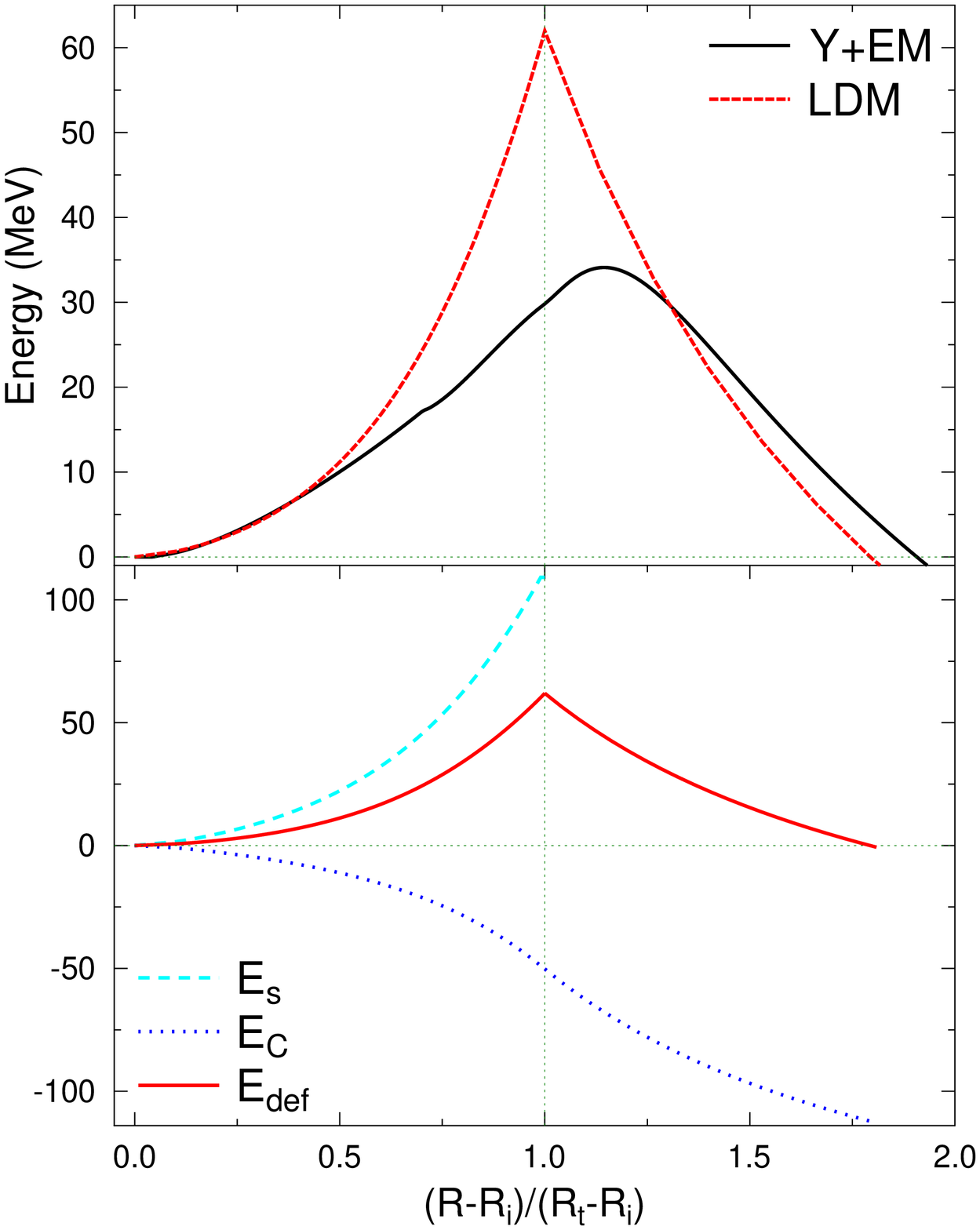} 
\caption{(Color online) TOP: potential barrier for emission of $^{34}\bar{Si}$
from $^{242}\bar{Cm}$ calculated within LDM (red) Y+EM (black). BOTTOM:
two main terms of the LDM barrier: surface energy (dashed line cyan)
and Coulomb energy (dotted line blue). 
\label{cmsi}}
\end{figure}

\begin{figure}[htb]
\centerline{\includegraphics[width=14cm]{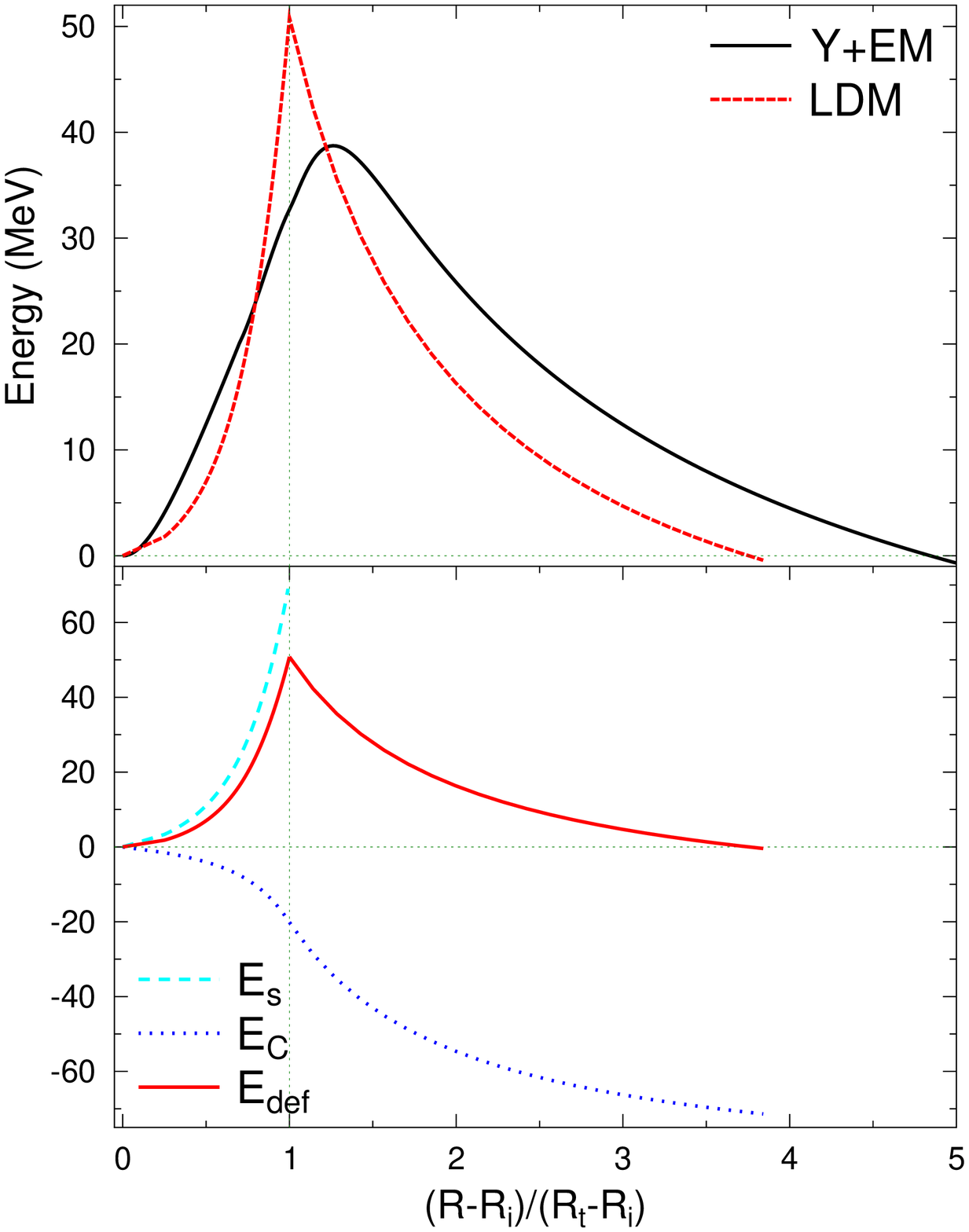}} 
\caption{(Color online)  TOP: potential barrier for emission of $^{14}\bar{C}$
from $^{250}\bar{Cf}$ calculated within LDM (red) Y+EM (black). BOTTOM:
two main terms of the LDM barrier: surface energy (dashed line cyan)
and Coulomb energy (dotted line blue). 
\label{cfc}}
\end{figure}

In figures \ref{cmsi} and \ref{cfc} (top panel) we present two examples of
potential barriers calculated within LDM and Y+EM for spontaneous emission
of $^{34}\bar{Si}$ from $^{242}\bar{Cm}$ and $^{14}\bar{C}$ radioactivity of
$^{250}\bar{Cf}$, respectively.  In both cases it is clear that within Y+EM
the strong interaction continues to act, as a proximity force even for
separated fragments, $R>R_t$, as long as the the tip separation distance
remains small enough; the interaction energy is maximum at certain distance
$R_m > R_t$.  For spherical fragments there is an analytical relationships
of interaction term:
\begin{equation}
E_{Y12} = -4\left ( \frac{a}{r_0} \right ) ^2 \sqrt {a_{21} a_{22}}
\left [ g_1 g_2 \left ( 4+\frac{R}{a} \right ) -g_2f_1 - g_1f_2
\right ] \frac{\exp (- R/a)}{R/a}
\end{equation}
\begin{equation}
g_k =  \frac{R_k}{a} \cosh \left ( \frac{R_k}{a}  \right ) - \sinh
\left ( \frac{R_k}{a} \right ) \; ; \; f_k = \left (\frac{R_k}{a} 
\right ) ^2 \sinh \left ( \frac{R_k}{a} \right )
\end{equation} 
where $a=0.68$ fm is the diffusivity parameter and $a_2 = a_s(1-\kappa
I^2)$, $a_s=21.18466$~MeV, $\kappa =2.345$.

The contribution of surface, $E_s$, and Coulomb energy, $E_C$, to the LDM
potential barrier is plotted at the bottom of figures \ref{cmsi} and
\ref{cfc}.  The potential barrier height is the result of adding an
increasing with separation distance surface energy up to the touching point
with a decreasing electrostatic energy up to infinity.

\section{Half-lives}

The experimental data on halflives against cluster radioactivity
\cite{bon07rrp,gug08jpcs}, $T_c$, and $\alpha$~decay, $T_\alpha$, are given
in Table~\ref{tab1}, together with $Q$-values, updated using the mass tables
published in 2012 \cite{wan12cpc}. Up to now there was not observed any
odd-odd cluster emitter.

\begin{table}[hbt] 
\caption{$Q$values in MeV and decimal logarithm of the half-lives in seconds
for the most probable CR and $\alpha$D of cluster emitters experimentally
observed.
\label{tab1}} 
\begin{center}
\begin{ruledtabular}
\begin{tabular}{c|c|ccc|cc}
Parent&Emitted&$Q_c$  &$\log_{10} T^{exp}_c(s)$&$\log_{10}
T^{ASAF}_c(s)$ &   $Q_\alpha$ &$\log_{10} T^{exp}_\alpha (s)$\\
\hline
$^{221}$Fr& $^{14}$C& 31.291& 14.52& 14.27&6.458& 2.55\\
$^{221}$Ra& $^{14}$C& 32.395& 13.39& 13.74&6.881& 1.90\\
$^{222}$Ra& $^{14}$C& 33.049& 11.01& 11.15&6.679& 1.58\\
$^{223}$Ra& $^{14}$C& 31.828& 15.19& 14.72&5.979& 5.99\\    
$^{224}$Ra& $^{14}$C& 30.534& 15.86& 15.93&5.789& 5.50\\
$^{226}$Ra& $^{14}$C& 28.196& 21.19& 20.98&4.870&10.70\\
$^{223}$Ac& $^{14}$C& 33.064& 12.96& 12.68&6.783& 2.48\\
$^{225}$Ac& $^{14}$C& 30.476& 17.28& 17.69&5.935& 6.23\\ \hline

$^{228}$Th& $^{20}$O& 44.724& 20.72& 21.72&5.520& 7.78\\ \hline

$^{231}$Pa& $^{23}$F& 51.860& 26.02& 25.52&5.149&11.47\\ \hline

$^{230}$U&  $^{22}$Ne&61.386& 19.57& 20.12&5.992& 6.26\\ \hline

$^{230}$Th& $^{24}$Ne&57.760& 24.61& 24.86&4.770&12.38\\
$^{231}$Pa& $^{24}$Ne&60.409& 23.23& 23.01&5.149&11.47\\
$^{232}$U&  $^{24}$Ne&62.309& 20.42& 20.37&5.413& 9.34\\
$^{233}$U&  $^{24}$Ne&60.485& 24.84& 24.97&4.909&12.78\\
$^{234}$U&  $^{24}$Ne&58.824& 25.92& 25.72&4.857&13.04\\
$^{235}$U&  $^{24}$Ne&57.362& 27.42& 29.97&4.678&16.57\\ \hline

$^{233}$U&  $^{25}$Ne&60.727& 24.84& 25.48&4.909&12.78\\
$^{235}$U&  $^{25}$Ne&57.706& 27.42& 30.38&4.678&16.57\\ \hline

$^{234}$U&  $^{26}$Ne&59.415& 25.92& 26.59&4.857&13.04\\ \hline

$^{234}$U&  $^{28}$Mg&74.109& 25.14& 25.34&4.857&13.04\\
$^{236}$Pu& $^{28}$Mg&79.668& 21.52& 20.55&5.867& 7.95\\
$^{238}$Pu& $^{28}$Mg&75.909& 25.70& 25.60&5.593& 9.44\\ \hline

$^{236}$U&  $^{30}$Mg&72.274& 27.58& 29.54&4.573&14.99\\
$^{238}$Pu& $^{30}$Mg&76.795& 25.70& 25.86&5.593& 9.44\\ \hline

$^{238}$Pu& $^{32}$Si&91.186& 25.27& 25.33&5.593& 9.44\\ \hline

$^{242}$Cm& $^{34}$Si&96.509& 23.15& 22.77&6.216& 7.15\\ 
\end{tabular}
\end{ruledtabular}
\end{center}
\end{table}

It is clear that $T_c >> T_\alpha$, hence cluster radioactivity of nuclei
with atomic numbers $Z=87-96$ is a rare phenomenon in a huge background of
$\alpha$~particles.  The measurements are in good agreement with predictions
within analytical superasymmetric fission (ASAF) model
\cite{p123adnd86,p160adnd91}.  Surprisingly, for some superheavy nuclei we
found \cite{p309prl11,p315prc12} comparable half-lives or even shorter $T_c
< T_\alpha$.

We expect that the same $Q$-values and half-lives will be observed in the
future for anti-cluster decay and anti-alpha decay of antimatter nuclei. 
Perhaps the easiest way to observe the decay modes of antimatter nuclei
would be to produce the lightest $\bar{\alpha}$ emiter, $^8\bar{Be}$, which
will be split in two $^4\bar{He}$ or two $\bar{\alpha}$ nuclei with a
half-life of about $81.9\cdot 10^{-18}$~s$=81.9$~as --- the same with that
of $^8Be \rightarrow \alpha + \alpha $ \cite{aud12cpc1}.

\begin{acknowledgments} 
This work was supported within the IDEI Programme under Contracts No. 
43/05.10.2011 and No.  42/05.10.2011 with UEFISCDI, and NUCLEU Programme,
Bucharest.  
\end{acknowledgments}


\newpage

\end{document}